\documentclass[a4paper,fleqn]{cas-dc}
\usepackage{graphicx}%
\usepackage{multirow}%
\usepackage{amsmath,amssymb,amsfonts}%
\usepackage{amsthm}%
\usepackage{mathrsfs}%
\usepackage[title]{appendix}%
\usepackage{xcolor}%
\usepackage{textcomp}%
\usepackage{manyfoot}%
\usepackage{booktabs}%
\usepackage{url}
\usepackage{algorithm}%
\usepackage{algorithmicx}%
\usepackage{algpseudocode}%
\usepackage{listings}%
\usepackage{makecell} 
\usepackage{float}
\usepackage{tcolorbox}
\usepackage{rotating}
\tcbuselibrary{listingsutf8}
\usepackage[numbers]{natbib}
\usepackage{ragged2e}
\justifying

\def\tsc#1{\csdef{#1}{\textsc{\lowercase{#1}}\xspace}}
\tsc{WGM}
\tsc{QE}
\tsc{EP}
\tsc{PMS}
\tsc{BEC}
\tsc{DE}

\begin{document}
\let\WriteBookmarks\relax
\def\floatpagepagefraction{1}
\def\textpagefraction{.001}
\shorttitle{Thyroid Scintigraphy Augmented}
\shortauthors{M. Sabouri et~al.}

\title [mode = title]{AI-Augmented Thyroid Scintigraphy for Robust Classification of Disease}               

\author[1,2]{Maziar Sabouri}[orcid=0000-0001-7525-8952]
\fnmark[1]
\ead{maziarsabouri@phas.ubc.ca}

\author[3]{Ghasem Hajianfar}
\fnmark[1]

\author[4]{Alireza Rafiei Sardouei}

\author[4]{Milad Yazdani}

\author[5]{Azin Asadzadeh}

\author[6]{Soroush Bagheri}

\author[7]{Mohsen Arabi}

\author[8]{Seyed Rasoul Zakavi}

\author[8]{Emran Askari}

\author[8]{Atena Aghaee}

\author[9]{Sam Wiseman}

\author[10, 11]{Dena Shahriari}

\author[3]{Habib Zaidi}

\author[1,2,10,12]{Arman Rahmim}[orcid=0000-0002-9980-2403]

\affiliation[1]{organization={Department of Physics \& Astronomy, University of British Columbia}, city={Vancouver}, country={Canada}}
\affiliation[2]{organization={Department of Basic and Translational Research, BC Cancer Research Institute}, city={Vancouver}, country={Canada}}
\affiliation[3]{organization={Division of Nuclear Medicine and Molecular Imaging, Department of Medical Imaging, Geneva University Hospital}, city={Geneva}, country={Switzerland}}
\affiliation[4]{organization={Department of Electrical and Computer Engineering, University of British Columbia}, city={Vancouver}, country={Canada}}
\affiliation[5]{organization={Department of Nuclear Medicine, 5Azar Hospital, Golestan University of Medical Sciences}, city={Gorgan}, country={Iran}}
\affiliation[6]{organization={Department of Medical Physics, Kashan University of Medical Sciences}, city={Kashan}, country={Iran}}
\affiliation[7]{organization={Department of Pathology and Radiology, School of Medicine, Alborz University of Medical Sciences}, city={Karaj}, country={Iran}}
\affiliation[8]{organization={Nuclear Medicine Research Center, Mashhad University of Medical Sciences}, city={Mashhad}, country={Iran}}
\affiliation[9]{organization={Department of Surgery, St. Paul's Hospital \&
University of British Columbia}, city={Vancouver}, country={Canada}}
\affiliation[10]{organization={School of Biomedical Engineering, University of British Columbia}, city={Vancouver}, country={Canada}}
\affiliation[11]{organization={Department of Orthopaedics, Faculty of Medicine, University of British Columbia}, city={Vancouver}, country={Canada}}
\affiliation[12]{organization={Department of Radiology, University of British Columbia}, city={Vancouver}, country={Canada}}

\fnmark[2]
\ead{arman.rahmim@ubc.ca}
\fntext[fn1]{Maziar Sabouri and Ghasem Hajianfar contributed equally to this work.}
\fntext[fn2]{Arman Rahmim is the corresponding author.}

\begin{abstract}
Thyroid scintigraphy is vital for diagnosing thyroid disorders, yet deep learning (DL) models in this domain often struggle with limited, imbalanced datasets. This study investigates the impact of three data augmentation strategies including Stable Diffusion (SD), Flow Matching (FM), and Conventional Augmentation (CA), on enhancing DL-based classification of disease. Anterior thyroid scintigraphy images from 2,954 patients across nine medical centers were classified into four categories: Diffuse Goiter (DG), Nodular Goiter (NG), Normal (NL), and Thyroiditis (TI). Data augmentation was performed using CA as well as various SD and FM models, creating 18 distinct scenarios. Each augmented dataset was used to train a ResNet18 DL-classifier. Model performance was assessed using class-wise and average precision, recall, F1-score, AUC, and image fidelity metrics (FID and KID). FM-based methods demonstrated top-tier performance, with the Original dataset combined with FM (O+FM) configuration achieving the highest micro, macro, and weighted F1-scores (0.78, 0.77, 0.78) and AUC values (0.95, 0.93, 0.94). While the O+FM+CA model also yielded excellent, balanced results, O+FM was statistically superior, indicating that high-fidelity generative augmentation can supersede conventional heuristics. FM also produced the most realistic images, achieving the lowest overall FID (0.66) and KID (0.83). Among the SD variants, SD1 combining image and prompt inputs was the most effective (macro F1: 0.76; FID: 4.17), showing that physician-generated prompts provide critical clinical context. Integrating FM and clinically-informed SD augmentation substantially improves thyroid scintigraphy classification, highlighting the importance of advanced generative models for robust training on limited datasets. The code is available at: \url{https://github.com/MaziarSabouri/Stable-Diffusion-Scintigraphy-Augmentation}
\end{abstract}


\begin{highlights}
\item Flow Matching-based augmentation leads to significantly improved thyroid scintigraphy classification compared to Stable Diffusion-based methods.
\item Flow Matching-based augmentation preserved key image features, enhancing model generalization and aiding consistent clinical diagnosis.
\item Clinically informed prompts fed to Stable Diffusion methods enhance image realism and significantly improve classification performance, underscoring the value of domain knowledge in generative augmentation. \\

\textbf{The paper was received by the \textit{Physica Medica} editorial office on 28 September 2025, revised on 9 May 2026, and officially accepted for publication on 13 June 2026.}

\end{highlights}


\maketitle

\section{Introduction}

Thyroid diseases are among the most common endocrine disorders, affecting millions of subjects worldwide \cite{pizzato2022epidemiological}, and their early and accurate diagnosis is essential for selecting appropriate treatments that lead to optimal patient outcomes. Physicians rely on a variety of imaging techniques, including ultrasound (US), computed tomography (CT), magnetic resonance imaging (MRI), and thyroid scintigraphy (gamma scan), along with laboratory tests, to evaluate and diagnose thyroid conditions \cite{bagheri2025impact}.

While each imaging modality serves a distinct clinical application, they also present specific limitations. Although US is widely used to evaluate nodular disease, its utility is dependent upon operator experience. CT and MRI can evaluate for structural characteristics, such as tracheal compression or substernal extension, but provide no functional information to assist with the diagnosis of conditions such as Graves’ disease \cite{calle2021imaging}. By contrast, thyroid scintigraphy (using the 99mTc-pertechnetate radiopharmaceutical) holds unique importance in the diagnostic workflow, as it provides crucial insight into both the structure and function of the thyroid gland. However, interpreting these images may be subjective, time-consuming, and prone to variability among experts. These challenges highlight the need for improved and objective automated approaches to enhance diagnostic accuracy and efficiency \cite{giovanella2019eanm}.

Artificial intelligence (AI) has shown significant potential in medicine, particularly in disease diagnosis, treatment guidance, and personalized patient care \cite{alowais2023revolutionizing}. However, a major challenge in developing deep learning (DL) models for medical applications is data scarcity. Large datasets are crucial for training robust models, yet collecting sufficient data can be difficult due to privacy concerns, high costs, and logistical constraints \cite{Shorten2019aug}. Conventional augmentation (CA) techniques, such as rotation, flipping, shifting, scaling, etc. help improve model generalization by creating variations of existing data \cite{islam2024systematic}. However, these methods alone are often insufficient to fully address data limitations, highlighting the need for more advanced strategies \cite{zhang2023diffusion}.

Recent studies have investigated advanced augmentation techniques to overcome the challenge posed by limited medical imaging datasets. While Generative Adversarial Networks (GANs) \cite{GAN_paper} and Variational Autoencoders (VAEs) \cite{VAE_paper} have shown success, diffusion-based models \cite{ho2020ddpm} demonstrate superior performance in image synthesis, producing highly realistic augmented images \cite{hajianfar2024stable,zhang2023diffusion,akrout2023diffusion}.

In this study, a comprehensive augmentation approach on thyroid scintigraphy images was implemented using diffusion-based algorithms, including Denoising Diffusion Probabilistic Models (DDPM) \cite{ho2020ddpm} and Flow-Matching (FM) \cite{lipman2023flow} to address the challenge of limited medical imaging data. To assess the effectiveness of the generated images, we incorporate them into the training process of a classification model and evaluate their performance on an external dataset.

In this work, our key contributions are as follows: 1) We explored novel application of diffusion-based models to augment thyroid scintigraphy images and address data scarcity. 2) We utilized physician reports to extract prompts used alongside input images in the Stable Diffusion models, guiding and enriching the image generation process. 3) We demonstrated that diffusion-based augmentation improves classification performance in thyroid scintigraphy imaging.

Paper architecture: The paper is organized as follows. Section \ref{Methodology} outlines the methods used. Section \ref{Experiments} covers the dataset, training setup for augmentation and classification, and evaluation metrics. Section \ref{Results} presents a detailed analysis of the results. Section \ref{Discussion} reviews related work and discusses our findings. Finally, Section \ref{Conclusion} summarizes the conclusions.

\section{Methodology} \label{Methodology}

\begin{figure*}[!htbp]
\centering
\includegraphics[width=0.7\linewidth]{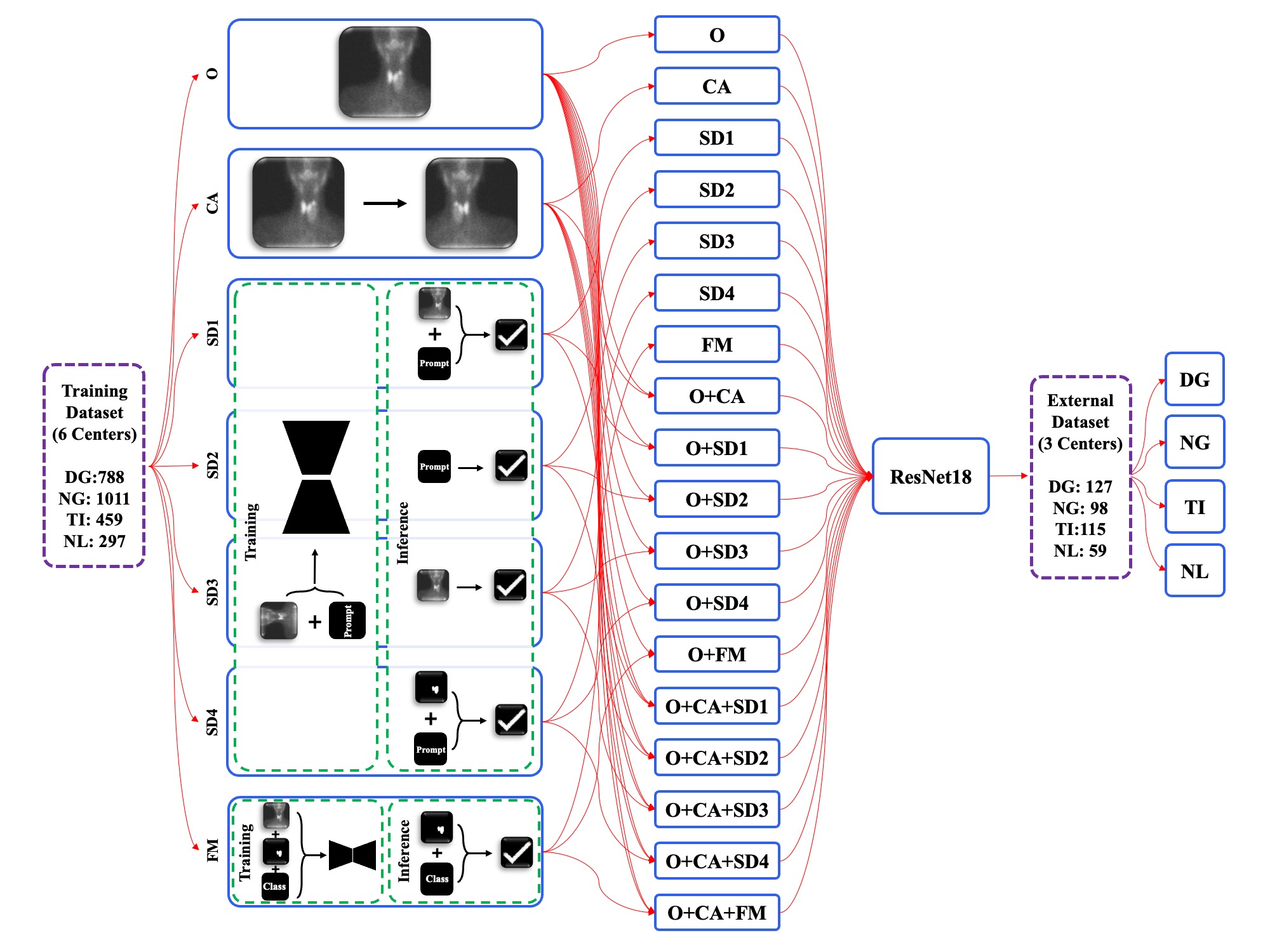}
\caption{Overview of the study workflow illustrating the dataset, augmentation strategies, and model training. (DG: Diffuse Goiter, NG: Nodular Goiter, NL: Normal, TI: Thyroiditis, O: Original, CA: Conventional Augmentation, SD: Stable Diffusion, FM: Flow Matching)}
\label{Fig: flowchart}
\end{figure*}

We aimed to find the best augmentation method to enhance the classification performance. Conventional methods \cite{conventional_augmentation} can generate cases that belong to completely different classes \cite{islam2024systematic}. Therefore, we need a method to learn the distribution of the dataset images and then draw samples from it. GANs\cite{GAN_paper}, Variational Autoencoders (VAEs) \cite{VAE_paper}, DDPMs \cite{ho2020ddpm} and FMs \cite{lipman2023flow} are some examples. Among these algorithms, DDPMs and FMs have exhibited superior performance \cite{DDPM_FM_better_GAN}. Hence, we consider these two approaches. The flowchart of the study is presented in Figure \ref{Fig: flowchart}.

\subsection{Stable Diffusion}
SD is a type of \textbf{Latent Diffusion Model (LDM)} \cite{StableDiffusion_paper}, which belongs to the DDPM family but operates in a lower-dimensional latent space to improve efficiency. Unlike standard DDPMs, which directly apply diffusion to raw image pixels, LDMs first encode the image into a compact latent representation using a pre-trained VAE. The diffusion process then operates in this latent space, making it computationally efficient while preserving high-quality image generation. For the diffusion process, there are two phases: the \textbf{Forward Process} and the \textbf{Reverse Process}.\\

\textbf{Forward Process}: Let the target image be denoted as $\mathbf{x}_0$. In the forward process, we gradually add Gaussian noise to the sample in a Markovian manner:
\[\mathbf{x}_t = \sqrt{1-\beta_t} \mathbf{x}_{t-1} + \sqrt{\beta_t} \mathbf{v}_t, \quad \mathbf{v}_t \sim \mathcal{N}(0, I)\]
The coefficients $\sqrt{1-\beta_t}$ and $\sqrt{\beta_t}$ control the transition, ensuring a gradual corruption of the data while maintaining variance stability. After a sufficient number of steps ($T$), the sample $\mathbf{x}_T$ follows a standard Gaussian distribution, i.e., 
\[\mathbf{x}_T \sim \mathcal{N}(0, I)\]
\textbf{Reverse Process}: To generate new samples, we approximate the reverse diffusion process. Starting from $\mathbf{x}_T \sim \mathcal{N}(0, I)$, we iteratively sample from the conditional distribution:
\[\mathbf{x}_{t-1} \sim p_\theta(\mathbf{x}_{t-1} | \mathbf{x}_t)\]
Since $q(\mathbf{x}_{t-1}|\mathbf{x}_t)$ is intractable, we assume it follows a Gaussian distribution and train a neural network (e.g., a U-Net) to estimate the necessary parameters for sampling. By iterating this denoising process from $T$ down to $0$, we obtain a generated sample $\hat{\mathbf{x}}_0$. \\

Since we are using the SD model, we can incorporate additional \textbf{conditioning information} into the reverse process. This is achieved by modifying the reverse process to be conditional on auxiliary inputs such as text prompts or images. The new conditional distribution is given by:
\[\mathbf{x}_{t-1} \sim p_\theta(\mathbf{x}_{t-1} | \mathbf{x}_t, C)\] 
where $C$ represents the selected condition, which can be $P$ (a prompt), $M$ (a mask), or 
$y$ (a given image). To enable conditioning, SD trains the noise prediction model $p_\theta$ to take both $x_t$ and $C$ as inputs, ensuring that the generated sample aligns with the provided condition. This makes text-to-image and mask-to-image generation possible within the SD framework. 

\subsection{Flow matching}
FM provides an alternative to diffusion models by directly learning a continuous-time velocity field that defines a near-optimal transport between the source and target distributions. Instead of progressively adding and then removing noise, FM defines a straight-line (or nearly straight) transformation between cases from the data distribution and a known prior. This makes sampling more efficient compared to traditional diffusion-based approaches. \\

Let $\mathbf{x}_0 \sim p_0(x)$ and $\mathbf{x}_1 \sim p_1(x)$ represent two distributions, where $\mathbf{x}_0$ is the source distribution (e.g., real data) and $\mathbf{x}_1$ is the target distribution (e.g., noise or another transformed version of the data). FM constructs a continuous interpolation between these two distributions as:
\[\mathbf{x}_t = t \mathbf{x}_1 + (1-t) \mathbf{x}_0, \quad t \in [0,1]\] 
This formulation defines a linear transport path from $\mathbf{x}_0$ to $\mathbf{x}_1$. The goal is to learn a velocity field $\mathbf{v}_\theta (\mathbf{x}_t,t)$ that describes the optimal transport direction at each time step. Ideally, this velocity field should satisfy:
\begin{equation}
v_\theta(\mathbf{x}_t, t) = \mathbf{x}_1 - \mathbf{x}_0 
\end{equation}
To ensure that the learned velocity field $\mathbf{v}_\theta (\mathbf{x}_t,t)$ correctly follows the transport direction, we minimize the FM loss: 
\begin{equation}
\mathcal{L} = \mathbb{E}_{\mathbf{x}_0, \mathbf{x}_1} \left[ | (\mathbf{x}_1 - \mathbf{x}_0) - v_\theta(\mathbf{x}_t, t) |_2^2 \right] 
\end{equation}
This objective encourages the model to approximate the optimal transport map under a quadratic cost, ensuring that the flow remains efficient and direct.

During inference, novel cases can be generated by solving the learned ordinary differential equation (ODE) defined by the velocity field: 
\begin{equation}
\frac{d\mathbf{x}_t}{dt} = v_\theta(\mathbf{x}_t, t) 
\end{equation}
This ODE governs the smooth transport from $\mathbf{x}_0$ to $\mathbf{x}_1$. Unlike diffusion models, which require many discretized steps for effective denoising, FM provides a single-step or low-step approximation to recover the target distribution, making it significantly more efficient in practice.

\subsection{ResNet18}
ResNet-18 \cite{he2015deepresiduallearningimage} is a deep convolutional neural network (CNN) designed to enable effective feature extraction while addressing the vanishing gradient problem through residual learning. Unlike traditional CNN architectures that rely solely on stacked convolutional layers, ResNet introduces skip connections, allowing gradient flow across layers and improving convergence during training. 

While attention-based architectures, such as Transformers \cite{vaswani2023attentionneed}, have gained popularity for complex datasets requiring long-range dependencies, our dataset does not exhibit such complexity. Instead, the patterns in our data can be effectively captured using local feature extraction mechanisms, making convolutional architectures a suitable choice. Given that ResNet-18 provides a strong balance between depth and computational efficiency, we achieve high performance without the need for more computationally expensive architectures. 

ResNet-18 consists of an initial convolutional layer, followed by four residual stages, and ends with global average pooling and a fully connected layer. Each residual block applies two 3×3 convolutions with identity shortcuts, ensuring efficient feature learning. The architecture follows:
\begin{center}
\scalebox{0.85}{$
\text{Conv(7×7,64)→MaxPool(3×3)→Residual Block×4→AvgPool→FC}
$}
\end{center}

\section{Experiments} \label{Experiments}

Table \ref{tab:data_summary} provides an overview of the studied cases collected from nine centers using eight different imaging systems. The dataset covers a broad age range with a mean age of 44.71 ± 17.66 years. The gender distribution includes 771 males (26\%) and 2,183 females (74\%). The cases have been classified into four categories: Diffuse Goiter (DG), Nodular Goiter (NG), Thyroiditis (TI), and Normal (NL), totaling 2,954 cases. Data usage for this study was approved by the Research Ethics Committee of Kashan University of Medical Sciences (IR.KAUMS.NUHEPM.REC.1403.022), and the research was conducted in accordance with the Declaration of Helsinki.

\begin{table*}[!htbp]
    \centering
    \caption{Data information and distribution across different centers.}
    \label{tab:data_summary}
    \renewcommand{\arraystretch}{1}
    \fontsize{10pt}{14pt}\selectfont
    \begin{tabular}{cccccll}
        \hline
        \textbf{Center} & \textbf{Age} & \textbf{M/F} & \textbf{DG/NG/TI/NL} & \textbf{Total} & \textbf{Manufacturer} & \textbf{Model} \\
        \hline
        \multicolumn{7}{c}{\textbf{Training Dataset}} \\
        \hline
        A      & NA                & 129/303     & 100/203/81/48  & 432   & ADAC         & GENESYS                  \\
        B      & 46.12 ± 15.26     & 77/166      & 63/110/47/23   & 243   & SIEMENS   & IP2 (ECAM1028)           \\
        C      & 45.08 ± 14.84     & 137/511     & 215/317/88/28  & 648   & SIEMENS   & IP2 (ECAM1028)           \\
        D      & 43.81 ± 22.26     & 150/448     & 145/199/134/120 & 598   & Mediso       & AnyScan                   \\
        E      & 47.04 ± 16.10     & 70/249      & 89/121/60/49   & 319   & SIEMENS   & IP1 (ECAM10482)          \\
        F      & 41.42 ± 14.75     & 91/224      & 176/61/49/29   & 315   & GE           & Discovery NM 630         \\
        \hline
        \multicolumn{7}{c}{\textbf{External Dataset}} \\
        \hline
        G      & 42.90 ± 12.00     & 21/50       & 20/21/24/6     & 71    & MiE          & SCINTRON                 \\
        H      & 46.78 ± 15.64     & 46/96       & 61/42/31/8     & 142   & SIEMENS   & Encore 2 (SYMBIA1071)    \\
        I      & 41.30 ± 15.20     & 50/136      & 46/35/60/45    & 186   & GE           & INFINIA                  \\
        \hline
        \textbf{Total} & \textbf{44.71 ± 17.66} & \textbf{771/2183} & \textbf{915/1109/574/356} & \textbf{2954} &  &  \\
        \hline
    \end{tabular}
    \textit{M: Male, F: Female, DG: Diffuse Goiter, NG: Nodular Goiter, TI: Thyroiditis, NL: Normal}
\end{table*}

\subsection{Preprocessing}

An experienced nuclear medicine physician manually segmented the thyroid region from the scintigraphy images using the manual contouring tool in ITK-Snap software \cite{yushkevich2006userguided}. Among the 2,954 images, 319 (all from center E) had a resolution of 256 × 256, while the rest were 128 × 128. Images and masks from center E were resampled to 128 × 128 using BSpline and nearest neighbor interpolation, respectively.

In this study, we use physicians' case reports for synthesizing images, so consistency is crucial due to varying styles and approaches across different centers. First, we analyzed all reports using GPT-4 Turbo \cite{openai2024chatgpt} and generated questions to extract the most relevant information. These questions were then reviewed and refined by an experienced nuclear medicine physician. Finally, we used the revised questions to gather consistent information using GPT-4 Turbo and create prompts under 77 tokens to feed SD (Code Snippet \ref{lst:Code}). Additionally, for each center, an experienced technician randomly reviewed 50 cases of the generated prompts.
\newpage

\begin{lstlisting}[caption={Structured information extraction from thyroid scan reports}, label={lst:Code}]
prompt = f"\ You are a medical AI specialized in nuclear medicine.
Your task is to analyze a thyroid scan report and extract structured information.
Given the following thyroid scan report: {report}
Answer the following questions in a structured JSON format:
    "Class": "DG, NG, TI, NL",
    "thyroid_function_classification": "Hypofunction / Hyperfunction / Normal / Indeterminate",
    "radiotracer_uptake_pattern": "Homogeneous / Inhomogeneous / Focal / Diffuse",
    "has_nodules": "Yes / No",
    "multinodular_goiter": "Yes / No",
    "nodule_type": "Hot / Cold / Not specified",
    "thyroid_size": "Normal / Mildly enlarged / Significantly enlarged / Atrophic",
    "diffuse_enlargement": "Yes / No"
\end{lstlisting}

\subsection{Training set-up}
We trained our models using data from six centers (A–F) and evaluated the classifier on an external dataset from three centers (G–I) for classification evaluation. Five augmentation methods were employed, including CA, three variations of SD, and FM. For each augmentation method, 1,000 images were generated per class. 

\subsubsection{Conventional augmentation}
We applied a randomized transformation pipeline that included rotation (±15°), horizontal flipping, translation (±10\%), scaling (0.8–1.2×), and Gaussian noise addition ($\sigma$ = 0.001–0.01). 

\subsubsection{Stable diffusion augmentation}
During training, each image was paired with its corresponding prompt and provided to the Stable Diffusion model. We used a fine-tuning setup with mixed precision (fp16), a resolution of 128×128, exponential moving average, gradient accumulation with four steps, and a batch size of 1, optimizing for 50,000 steps at a learning rate of 1e-5. For inference, the iterative denoising process was set to 50 steps, and we evaluated four approaches: 1) image and prompt to image (SD1), 2) prompt to image (SD2), 3) image to image (SD3), and 4) mask and prompt to image (SD4).

\subsubsection{Flow matching augmentation}
For FM, the model was optimized using the Adam optimizer with a learning rate of 1e-4 for 200 epochs. Our approach leverages FM with optimal transport to align predicted data flows with the target distribution. Specifically, the continuous-time velocity field is parameterized using a UNet-based neural network \cite{DBLP:journals/corr/abs-2112-10752} incorporating flash attention to improve memory efficiency. Class conditioning is achieved by incorporating a one-hot encoded vector via cross-attention, while mask conditioning is implemented by introducing an additional UNet to encode the mask. The decoder of this secondary network employs zero convolutions, and skip connections are established between its decoder and the main velocity-predicting network to incorporate the mask-based guidance. During inference, novel samples are generated from Gaussian noise by solving the governing differential equation (ODE) using only 10 integration steps, which maps a nearly straight-line transformation between the source and target distributions, thereby requiring significantly fewer steps than standard iterative diffusion processes. We assessed guided generation using a combination of mask and class conditioning. The complete architectural details of the FM model, including the specific mechanisms for class and mask conditioning utilized in this study, were originally developed and comprehensively described in \cite{yazdani2026flow}.
This resulted in 18 distinct training strategies for the ResNet18 classifier: 1) O, 2) CA, 3) SD1, 4) SD2, 5) SD3, 6) SD4, 7) FM, 8) O+CA, 9) O+SD1, 10) O+SD2, 11) O+SD3, 12) O+SD4, 12) O+FM, 14) O+CA+SD1, 15) O+CA+SD2, 16) O+CA+SD3, and 17) O+CA+SD4, 18) O+CA+FM. 

\subsubsection{ResNet18}
We trained a ResNet18 classifier, initializing it with ImageNet1K pre-trained weights. The first convolutional layer was modified to retain a 3×3 kernel, and the fully connected layer was replaced with a dropout layer (0.2) followed by a linear layer matching the number of classes. The model was trained for 300 epochs using the Adam optimizer (learning rate = 1e-4, weight decay = 1e-5) with a cross-entropy loss function. A learning rate scheduler (ReduceLROnPlateau) was applied, reducing the rate by a factor of 0.8 if validation loss plateaued for 10 epochs, with a minimum learning rate of 2e-5. Furthermore, the training and validation split was set at a 9:1 ratio with stratification.

\subsection{Evaluation metrics}

\subsubsection{Augmentation metrics}
Fréchet Inception Distance (FID) \cite{heusel2018} and Kernel Inception Distance (KID) \cite{binkowski2021} metrics have been used for class-wise and overall comparisons between generated and original images. In both cases, 200 images per class were randomly selected from each dataset for evaluation. 

\subsubsection{Classification metrics}
The classification performance was evaluated on an external dataset using metrics, including precision, recall, F1-score, accuracy, and the area under the Receiver Operating Characteristic curve (ROC AUC). Given that it was a multiclass task, we applied various averaging techniques encompassing micro, macro, and weighted to provide a thorough evaluation across all classes. 

\subsubsection{Statistical method}
We compared different strategies using bootstrapping with 1,000 repetitions and sampling with replacement. Accuracy distributions were analyzed, and pairwise comparisons were made using the Wilcoxon rank sum test. To control for the false discovery rate across multiple comparisons, p-values were adjusted using the Benjamini-Hochberg procedure, with adjusted p-values < 0.05 considered statistically significant.

\subsubsection{GradCam}
We used Gradient-weighted Class Activation Mapping (Grad-CAM) on the trained ResNet18 model to generate heatmaps, visualizing the image regions that influenced the classifier’s decisions.

\section{Results} \label{Results}
\begin{figure*}[!htbp]
\centering
\rotatebox{0}{\includegraphics[width=0.8\linewidth]{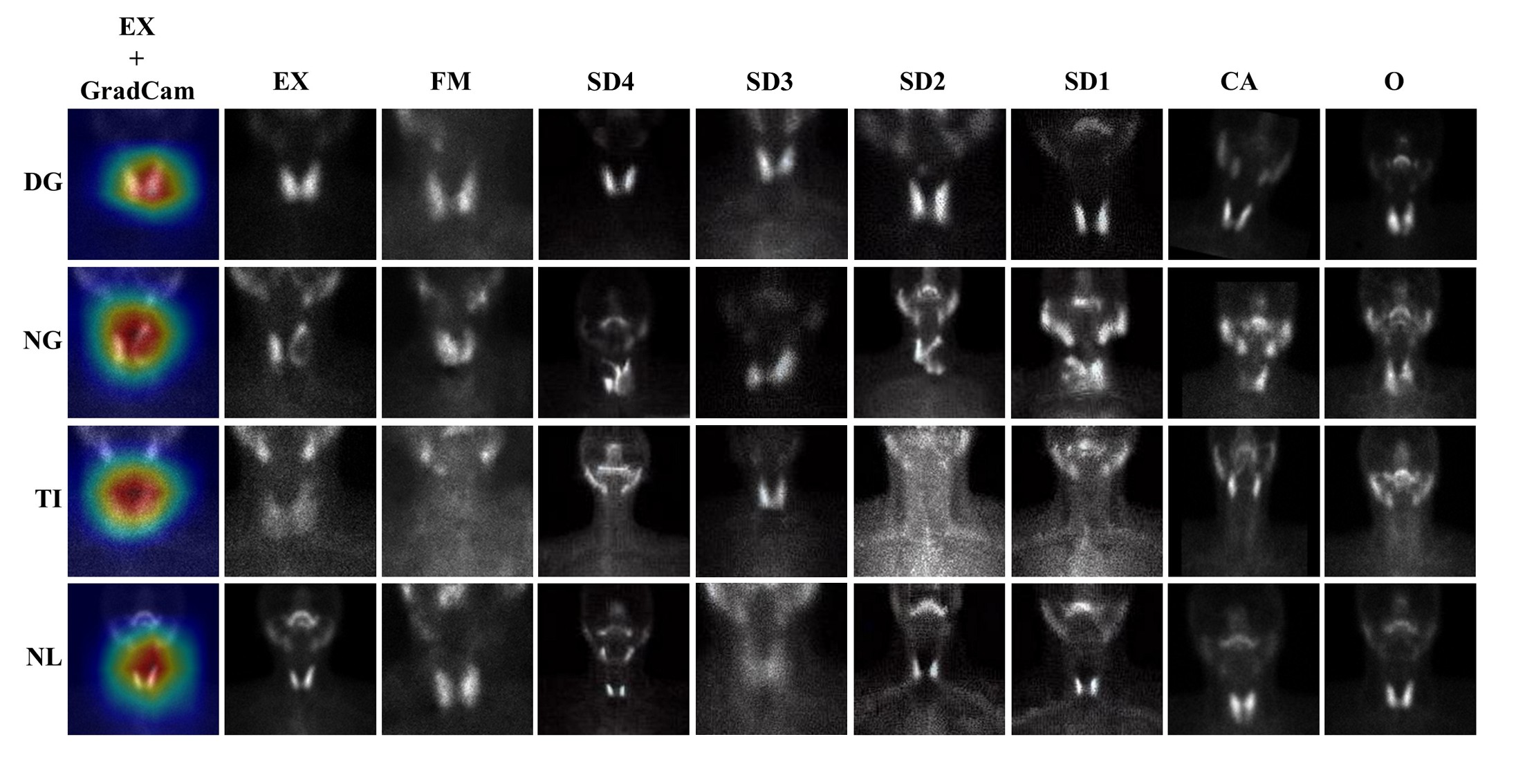}}
\caption{Examples of original and augmented images for each class using different methods. Grad-CAM visualizations from the O+FM model are also shown on external dataset samples, highlighting the model’s focus during prediction. (DG: Diffuse Goiter, NG: Nodular Goiter, NL: Normal, TI: Thyroiditis, O: Original, CA: Conventional Augmentation, SD: Stable Diffusion, FM: Flow Matching, EX: External)}
\label{Fig: samples}
\end{figure*}

Figure \ref{Fig: samples} presents one sample per class from the original dataset alongside examples generated by different augmentation methods used in this study. Additionally, it includes samples from the external dataset with their corresponding Grad-CAM visualizations using the O+FM model, highlighting the model's focus during prediction.

Table \ref{tab:merged_performance_metrics} reports class-wise and different averaging of precision, recall, F1-score, and AUC for each model variant. Among all configurations, both O+FM and O+FM+CA achieved top-tier performance, significantly outperforming the baseline and all SD-based methods. However, O+FM demonstrated a distinct advantage, yielding the highest micro, macro, and weighted F1-scores (0.78, 0.77, and 0.78, respectively) and the strongest AUC values (0.95, 0.93, 0.94). While O+FM+CA also produced excellent results with a highly balanced class-wise performance, the addition of Conventional Augmentation (CA) resulted in slightly lower average metrics compared to O+FM alone. This indicates that FM generates highly representative and robust samples on its own. Furthermore, FM alone already provides strong gains over O alone, especially in the challenging NL and TI classes, raising NL F1 from 0.39 to 0.51.

Among the SD-based methods SD1 shows the strongest standalone performance among SD variants, with solid macro F1 (0.75) and consistent gains across all classes. When combined with O (O+SD1), it improves macro F1 from 0.69 (O) to 0.72. SD2 is competitive when combined with O, with O+SD2 achieving F1 of 0.72/0.71/0.54/0.93 and macro AUC of 0.92. However, its standalone version (SD2) performs poorly in NL and TI classes. SD3 underperforms across the board. Its class-wise F1 scores (0.51/0.29/0.27/0.59) and overall metrics (macro F1: 0.42; AUC: 0.67) suggest limited utility in isolation. SD4 achieves high precision in the TI class (0.58), but extremely low recall (0.35 in NG) leads to imbalanced performance. Despite strong DG precision (0.88), its macro F1 (0.46) and micro accuracy (0.59) are among the lowest.

\begin{sidewaystable*}[!htbp]
    \centering
    \caption{Classification model performance metrics by class and average metrics for each augmentation method (*$\times 10^{-2}$).}
    \label{tab:merged_performance_metrics}
    \fontsize{8pt}{10pt}\selectfont
    \rotatebox{0}{
        \resizebox{\textheight}{!}{
    \begin{tabular}{@{}lccc|cccc@{}}
        \toprule
        \textbf{Method} & 
        \makecell{\textbf{Precision} \\ (DG/NG/NL/TI)} & 
        \makecell{\textbf{Recall} \\ (DG/NG/NL/TI)} & 
        \makecell{\textbf{F1-score} \\ (DG/NG/NL/TI)} & 
        \makecell{\textbf{AUC} \\ (mic/mac/wei)} & 
        \makecell{\textbf{Precision} \\ (mic/mac/wei)} & 
        \makecell{\textbf{Recall} \\ (mic/mac/wei)} & 
        \makecell{\textbf{F1-score} \\ (mic/mac/wei)} \\
        \midrule
        O & 70/65/53/96 & 72/86/31/91 & 71/74/39/93 & 92/91/91 & 73/71/72 & 73/70/73 & 73/69/71 \\
        CA & 75/70/55/92 & 62/77/62/97 & 68/73/58/95 & 93/92/92 & 74/73/75 & 74/74/74 & 74/74/74 \\
        SD1 & 80/74/53/92 & 72/71/59/98 & 76/72/56/95 & 93/92/92 & 76/74/76 & 76/75/76 & 76/75/76 \\
        SD2 & 76/57/45/95 & 70/68/53/74 & 73/62/49/83 & 88/87/88 & 67/68/70 & 67/66/67 & 67/67/68 \\
        SD3 & 55/27/32/54 & 48/32/23/66 & 51/29/27/59 & 68/67/67 & 44/42/44 & 44/42/44 & 44/42/43 \\
        SD4 & 88/66/42/58 & 44/35/64/100 & 58/46/50/74 & 82/86/86 & 59/63/67 & 59/61/59 & 59/57/58 \\
        FM & 70/54/48/96 & 67/64/54/76 & 69/59/51/85 & 89/88/88 & 66/67/69 & 66/65/66 & 66/66/67 \\
        \midrule
        O+CA & 73/59/61/99 & 72/83/41/89 & 72/69/49/94 & 92/92/92 & 73/73/74 & 73/71/73 & 73/71/73 \\
        O+SD1 & 74/67/55/96 & 70/85/46/89 & 72/75/50/92 & 92/92/92 & 74/73/74 & 74/73/74 & 74/72/74 \\
        O+SD2 & 78/60/61/98 & 67/87/49/89 & 72/71/54/93 & 92/92/92 & 74/74/76 & 74/73/74 & 74/73/74 \\
        O+SD3 & 72/62/57/93 & 71/83/39/87 & 72/71/46/90 & 90/90/90 & 72/71/72 & 72/70/72 & 72/70/71 \\
        O+SD4 & 76/62/59/95 & 66/87/47/89 & 71/73/53/92 & 92/92/92 & 73/73/74 & 73/72/73 & 73/72/73 \\
        \textbf{O+FM} & \textbf{74/74/68/93} & \textbf{81/76/54/95} & \textbf{77/75/60/94} & \textbf{95/93/94} & \textbf{78/77/78} & \textbf{78/76/78} & \textbf{78/77/78} \\
        O+SD1+CA & 75/64/54/96 & 66/82/50/91 & 70/72/52/93 & 93/92/93 & 73/72/74 & 73/72/73 & 73/72/73 \\
        O+SD2+CA & 74/65/62/94 & 66/82/54/91 & 70/73/58/92 & 92/92/92 & 74/73/74 & 74/73/74 & 74/73/74 \\
        O+SD3+CA & 77/63/60/91 & 65/78/58/90 & 71/70/59/90 & 92/91/91 & 73/73/74 & 73/73/73 & 73/72/73 \\
        O+SD4+CA & 72/63/58/95 & 63/82/53/90 & 67/72/55/92 & 92/91/92 & 72/72/73 & 72/72/72 & 72/72/72 \\
        \textbf{O+FM+CA} & \textbf{79/68/63/97} & \textbf{73/84/58/90} & \textbf{76/75/61/93} & \textbf{93/92/92} & \textbf{77/77/78} & \textbf{77/76/77} & \textbf{77/76/77} \\
        \bottomrule
        \end{tabular}

            }
}
\end{sidewaystable*}

\begin{figure}[!htbp]
\centering
\includegraphics[width=1\linewidth]{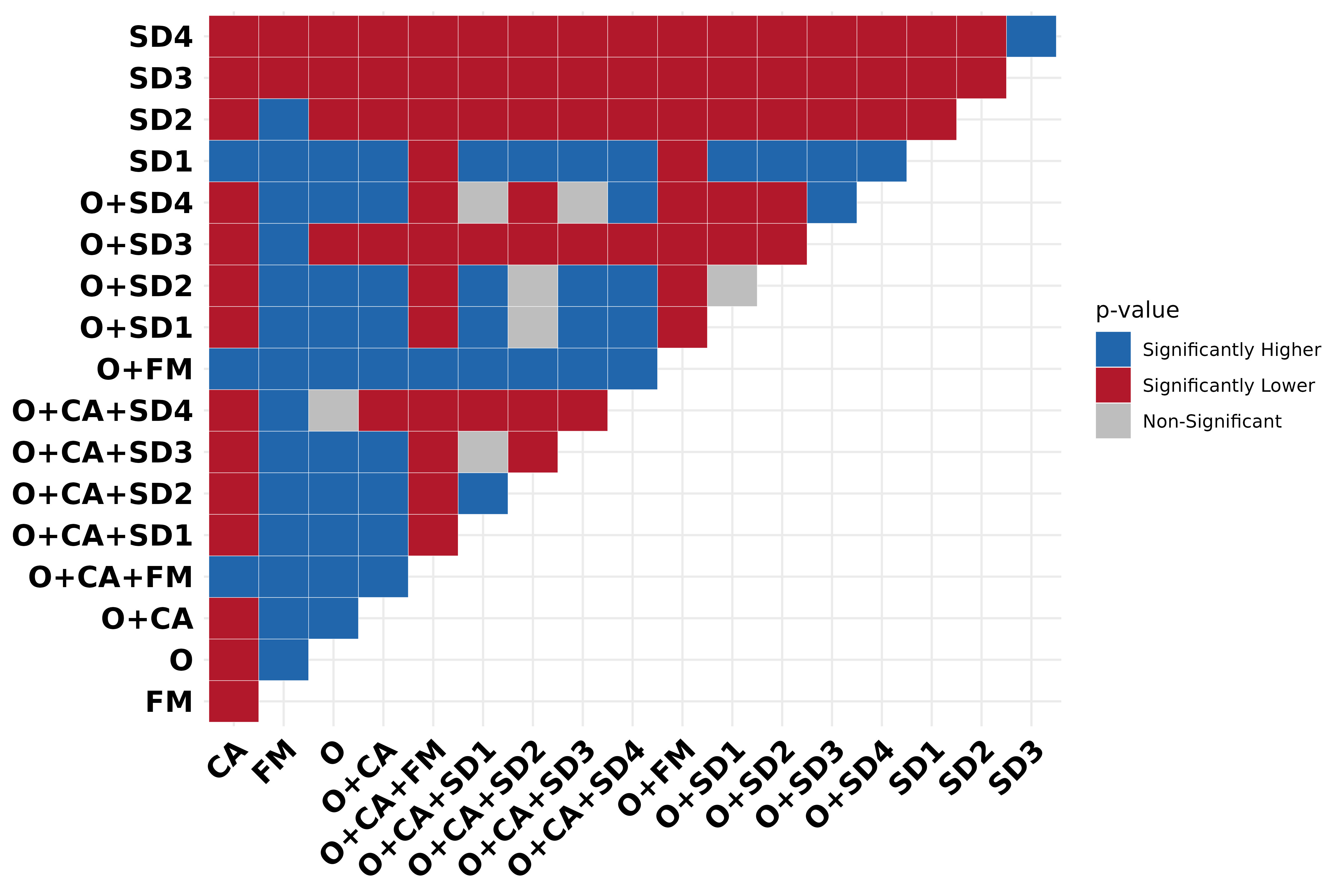}
\caption{Pairwise model comparison via the Wilcoxon signed-rank test. Each cell compares two models: Blue (row model significantly better), Red (worse), and Gray (no significant difference). (DG: Diffuse Goiter, NG: Nodular Goiter, NL: Normal, TI: Thyroiditis, O: Original, CA: Conventional Augmentation, SD: Stable Diffusion, FM: Flow Matching)}
\label{Fig: wilcoxon}
\end{figure}

Figure \ref{Fig: wilcoxon} presents a pairwise comparison of different models using the Wilcoxon rank sum test, highlighting significant performance differences. The results demonstrate that both O+FM and its extended variant (O+FM+CA) achieve consistently superior performance compared to almost all other models. Notably, within this top tier, O+FM significantly outperforms O+FM+CA, confirming it as the optimal standalone augmentation strategy. Furthermore, SD3 and SD4 exhibit the weakest performance, being significantly outperformed by almost all other models. Additionally, O+SD1, O+SD2, and their CA-augmented counterparts exhibit stronger results than SD2, SD3, and SD4 alone, further validating the impact of structured generation methods.

Table \ref{tab:aug_comparison} highlights clear differences in generative quality across augmentation methods, based on FID and KID scores. FM achieves the best overall performance, with the lowest FID (0.66) and KID (0.83), indicating that its generated images closely align with the real data distribution across all classes. Among the SD-based methods, SD1 and SD2 show moderate performance. SD2 achieves a lower overall FID (3.88) than SD1 (4.17), but its KID is slightly higher (2.61 vs. 4.99), and its performance on the TI class is notably poor (KID of 30.02). SD1 is more stable across classes, with no extreme outliers, though its FID and KID are consistently higher than FM. SD3 outperforms SD4 overall, with a lower FID (2.77 vs. 17.99) and KID (4.66 vs. 33.59). However, SD3 still lags behind FM, particularly in the TI class (FID: 4.15 vs. FM’s 1.97). SD4 performs the worst by a significant margin, especially in the NG, NL, and TI classes, indicating high visual artifacts and weak alignment with real data. For example, its TI KID reaches 50.50, far exceeding all others. Across all classes, FM consistently produces the lowest FID and KID, with the best performance in NG (FID: 0.74) and DG (FID: 0.96), and strong results even in the more challenging NL and TI categories. In contrast, the SD methods vary widely, with no single SD variant performing best across all classes. This underscores FM’s robustness and the instability of diffusion-based augmentation without careful conditioning.

\begin{table*}[!htbp]
    \centering
    \caption{Comparison of FID and KID scores across synthetic data generation methods.}
    \label{tab:aug_comparison}
    \fontsize{8pt}{10pt}\selectfont
    \begin{tabular}{@{}l*{10}{c}@{}}
        \toprule
        \textbf{Class} & \multicolumn{2}{c}{\textbf{SD1}} & \multicolumn{2}{c}{\textbf{SD2}} & \multicolumn{2}{c}{\textbf{SD3}} & \multicolumn{2}{c}{\textbf{SD4}} & \multicolumn{2}{c}{\textbf{FM}} \\
        \cmidrule(lr){2-3} \cmidrule(lr){4-5} \cmidrule(lr){6-7} \cmidrule(lr){8-9} \cmidrule(lr){10-11}
        & \textbf{FID} $\downarrow$ & \textbf{KID} $\downarrow$ 
        & \textbf{FID} $\downarrow$ & \textbf{KID} $\downarrow$ 
        & \textbf{FID} $\downarrow$ & \textbf{KID} $\downarrow$ 
        & \textbf{FID} $\downarrow$ & \textbf{KID} $\downarrow$ 
        & \textbf{FID} $\downarrow$ & \textbf{KID} $\downarrow$ \\
        \midrule
        DG  & 4.75  & 6.77  & 2.94  & 2.31  & 1.56  & 1.94  & 10.30 & 15.40 & \textbf{0.96}  & \textbf{0.80} \\
        NG  & 6.22  & 10.28 & 3.27  & 2.66  & 2.95  & 4.41  & 22.07 & 38.34 & \textbf{0.74}  & \textbf{0.95} \\
        NL  & 4.56  & 7.14  & 1.75  & \textbf{2.05}  & 3.39  & 5.68  & 18.68 & 36.74 & \textbf{0.85}  & 2.08 \\
        TI  & 2.73  & \textbf{3.01}  & 9.24  & 30.02 & 4.15  & 8.11  & 22.37 & 50.50 & \textbf{1.97}  & 3.39 \\
        \midrule
        \textbf{Overall} 
             & \textbf{4.17} & \textbf{4.99} 
             & \textbf{3.88} & \textbf{2.61} 
             & \textbf{2.77} & \textbf{4.66} 
             & \textbf{17.99} & \textbf{33.59} 
             & \textbf{0.66} & \textbf{0.83} \\
        \bottomrule
    \end{tabular}

    \vspace{0.3em}
    \footnotesize\textit{Standard deviations are less than 1e-5 for all measurements.}
\end{table*}

\section{Discussion} \label{Discussion}

Several studies have used diffusion-based models for data augmentation, leading to improved performance in classification tasks. However, only a few have incorporated physician reports as prompts to guide the diffusion process, which they reported to further enhance results. While all these works employed diffusion models, none explored the use of FM, which offers potential advantages in terms of efficiency and image quality. 

Zhang et al. \cite{zhang2023diffusion} investigated SinDDM \cite{kulikov2023sinddm}, a single-image denoising diffusion model, to augment lung ultrasound data. They also introduced FewDDM, an extension trained on limited samples, which outperformed single-image GANs in generating high-quality synthetic images. Augmenting with SinDDM notably improved pathology classification, especially for minority classes. Despite generating less detailed images, FewDDM surpassed SinDDM and SinGAN in downstream performance by capturing local structural variations. The study highlighted that combining synthetic and CA techniques yielded the best classification results. 

Hajianfar et al. \cite{hajianfar2024stable} investigated the effectiveness of SD \cite{StableDiffusion_paper} as an advanced augmentation method in enhancing deep learning models for classifying scintigraphic thyroid images. They used reports from physicians without specific cleaning as prompts in the augmentation process. The SD, O, and CA were used to train a ResNet101V2 classifier in different scenarios. The results demonstrated that models trained with synthetic data achieved consistently better performance. 

Balla et al. \cite{balla2024diffusion} explored strategies to address data scarcity in musculoskeletal US for osteoarthritis detection. They used CA with diffusion-based image synthesis, and the results showed that synthetic images generated through diffusion models retained anatomical fidelity and improved model generalization diagnostic accuracy, while CA sometimes hindered performance, highlighting the potential of using synthetic images. 

Akrout et al. \cite{akrout2023diffusion} advances data augmentation using text-to-image diffusion models to enhance a macroscopic skin disease dataset. By using text prompts, they gain fine-grained control over the image generation process. The results show that this generative augmentation approach maintains classification accuracy even when trained on a fully synthetic dataset. 

FM \cite{lipman2023flow}, as a novel, more robust, and memory-efficient method for image synthesis has shown superiority over GANs and DDPMs. While it has not been widely used in the medical domain, it has demonstrated clear advantages. 

In this study, we employed a variety of strategies for advanced augmentation. Specifically, we used image masks as conditions for both DDPM and FM, and we incorporated physicians' reports as prompts for DDPM to maximize the available information for augmentation. Using FM, our objective was to improve both the efficiency and quality of image synthesis, making it a key component of our approach. 

The comparison between O+FM and O+FM+CA reveals important insights into the nature of generative versus heuristic augmentation. Both methods achieved the highest overall performance in our study, underscoring the power of FM in handling data scarcity. However, our statistical analysis shows that O+FM significantly outperforms O+FM+CA in average F1-scores and AUC. This suggests that FM effectively learns and samples from the true underlying data distribution, creating highly realistic, anatomically and pathologically accurate variations. Applying conventional augmentation (such as random rotations, scaling, and noise) on top of this distribution does not yield additional overall benefits and may even introduce slight noise that marginally reduces the model's confidence. Therefore, high-fidelity generative models like FM are highly effective on their own.

Among the SD-based models, SD1 which uses both the image and prompt during inference, consistently outperforms SD2, SD3, and SD4, indicating that combining visual and textual information leads to higher-quality synthetic data and improved classification performance. This advantage may stem from two factors: either the SD model inherently performs better when both image and prompt are available, especially the ones were existed in the training process, or the prompts themselves, derived from physician reports, carry clinically valuable context that enhances generation. In this study, the latter appears especially plausible, as these prompts are grounded in real diagnostic language specific to thyroid scintigraphy, potentially guiding the model to produce more realistic and relevant variations. 

SD3, which performs image-to-image translation without textual input, ranks below SD1 and SD2. While it benefits from the visual structure of the original image, the lack of prompt input may limit its ability to generate semantically diverse or diagnostically meaningful augmentations. SD2, which performs prompt-only generation, achieves competitive overall FID but shows instability across classes, particularly with high KID in the TI category. This suggests that, while prompts can guide generation, relying on them alone may not provide enough structural information, especially for visually complex classes. SD4 performs the worst overall. It uses prompt and mask inputs during inference, but since masks were not present during training, the model lacks the capacity to meaningfully interpret them. As a result, the generated outputs are less coherent and lead to poor downstream performance. 

Compared to the study by Hajianfar et al. \cite{hajianfar2024stable}, which evaluated multiple augmentation strategies including Stable Diffusion (similar to SD1 in our study), CA, and their combinations, our work introduces a more advanced augmentation pipeline by incorporating FM, a technique not examined in their framework, along with three additional SD-based model variants. Both studies support the benefit of combining synthetic and real data over using original images alone. However, our results demonstrate that FM consistently generates higher-fidelity images, as evidenced by superior FID and KID scores, which correlate with improved micro- and macro-averaged F1-scores. While they identified the SD1+O as the most effective approach, our FM-based methods, particularly O+FM and O+FM+CA, achieved better class-wise balance and generalization, supported by statistical analysis and Grad-CAM visualizations on external data. Furthermore, implementing SD3 allows us to evaluate the added value of prompts, which has not been done in their study.

The strong correlation between image quality metrics (FID and KID) and classification performance highlights the clinical importance of generating realistic synthetic data. FM consistently achieved the lowest FID and KID scores, reflecting its ability to produce high-fidelity images that enhance model generalization and reliability. In contrast, SD3 and SD4’s poor image quality was associated with degraded classification performance, illustrating that not all augmentation methods are beneficial. High-quality, distribution-aligned synthetic data is essential to avoid spurious correlations and ensure model trustworthiness.  

Beyond image quality and classification accuracy, it is critical to contextualize the practical applicability of these generative methods in terms of computational cost and efficiency. Generative augmentation introduces a computational overhead, primarily during the training and sampling phases. Stable Diffusion, while operating efficiently in a compressed latent space, still requires a lengthy iterative denoising process during inference to generate high-quality images, which can be computationally expensive and time-consuming. Quantitatively, generating images at the 128×128 resolution using our Stable Diffusion pipeline required 50 iterative denoising steps, taking an average of 1.25 seconds per image on an NVIDIA GeForce RTX 4090 GPU. In contrast, FM directly models a continuous-time velocity field. By defining a nearly straight-line transport between distributions, FM allows the use of ODE solvers that require significantly fewer integration steps during inference. Specifically, the Flow Matching model requires only 10 integration steps to achieve superior image fidelity, drastically reducing the generation time to 0.19 seconds per image on the same hardware. This results in faster generation times and a reduced memory footprint during sampling without compromising image fidelity. Consequently, the FM approach offers a superior efficiency trade-off, making it highly practical for deploying scalable, high-throughput data augmentation pipelines in resource-constrained clinical environments \cite{yazdani2026flow}.

Furthermore, Grad-CAM visualizations on external data confirm that models trained with FM capture both the anatomical location and the functional (physiological) status of the thyroid gland, supporting interpretability and clinical reliability. As demonstrated in Figure \ref{Fig: samples}, the generated heatmaps accurately localize to the anatomical position of the thyroid gland across all four classes (DG, NG, TI, and NL). Crucially, the model demonstrates robust anatomo-physiological awareness even in diagnostically challenging scenarios such as TI. In TI cases, the physiological uptake of the radiotracer is often so low that the gland is barely visible; however, the model still successfully identifies and attends to the correct anatomical location to make its decision based on this lack of functional uptake. This consistent regional focus confirms that the FM augmentation strategy effectively trains the classifier to understand true anatomo-physiological features rather than relying on spurious background artifacts.

Regarding the potential dependency of the classification results on specific imaging hardware (manufacturer and model) or localized acquisition parameters, it is important to note that these variables are inherently collinear with the clinical centers in our multi-center dataset. However, the robustness of models trained on this data foundation against such scanner-specific variations has been previously established. In our prior study utilizing a Leave-One-Center-Out-Cross-Validation (LOCOCV) scheme on this dataset \cite{sabouri2024thyroidiomics}, classification models demonstrated highly stable performance across distinct clinical environments, yielding an accuracy of 0.76 $\pm$ 0.04 and a ROC AUC of 0.92 $\pm$ 0.02. This low variance across the LOCOCV folds confirms that the underlying features driving the classification are intrinsically tied to thyroid pathology rather than being artifacts of specific scanner manufacturers, models, or institutional imaging protocols. Consequently, by augmenting this already diverse dataset with high-fidelity FM, our current model inherits and further builds upon this robust generalizability.

It is worth noting that our experimental pipeline relied on a single classifier architecture, ResNet-18. While exploring higher-capacity backbones could yield higher absolute accuracy, the primary focus of this study was the comparative evaluation of generative augmentation strategies. Based on our prior work utilizing heavier architectures like ResNet101V2 for scintigraphy classification \cite{hajianfar2024stable}, we observed that the relative performance trends across different augmentation methods remain consistent regardless of the underlying classifier backbone. Given that our study necessitated training 18 distinct models, ResNet-18 was selected as the optimal evaluator, providing a necessary balance between robust feature extraction and the computational efficiency required for extensive comparative testing.

At last, it should be mentioned that the data in this study is limited to a single ethnic group; to ensure clinical applicability, further validation on more diverse datasets is necessary. Incorporating laboratory test results (e.g., TSH, free T3, free T4, and thyroid autoantibodies) and US images commonly used in these patients alongside nuclear medicine images could potentially enhance classification performance. Moreover, the use of reports from both imaging modalities, namely US and nuclear medicine, to generate prompts for SD-based models may further improve the augmentation results.

\section{Conclusion} \label{Conclusion}
In this study, we explored the impact of different augmentation strategies, including SD-, FM-, and CA-based, on thyroid classification using ResNet18. Our findings demonstrated that FM-based augmentation, particularly O+FM and O+FM+CA, consistently led to superior performance compared to all SD-based approaches. Between the top two configurations, O+FM showed a statistically significant advantage, achieving the highest overall accuracy, F1-scores, and AUC. This indicates that high-fidelity generative augmentation using FM is robust and highly effective on its own, precluding the strict need for conventional image transformations. 

Statistical significance testing using the Wilcoxon method reinforced these results, highlighting the effectiveness of FM in improving model generalization. FM ensures a smoother and more controlled transformation of image distributions, preserving essential structural details and intensity variations critical for classification. In contrast, SD-based models, especially those relying on image-only (SD3) or masked inputs (SD4), may introduce artifacts or inconsistencies that can mislead the classifier. This enhanced realism in FM-generated images leads to better feature representation learning, improving classification performance. At the same time, it is important to acknowledge that the dataset evaluated in this study is demographically and geographically limited. Future research must validate these generative augmentation strategies and classification models on broader, multi-ethnic global cohorts to ensure full generalizability and equitable clinical applicability across diverse patient populations.

Overall, this work presents a methodology enabling more objective and consistent thyroid scintigraphy analysis, and therefore diagnosis, with the potential to assist nuclear medicine physicians and to help improve patient outcomes. 

\section*{Statements and Declarations}

\subsection*{Funding}
This work was supported by the Natural Sciences and Engineering Research Council of Canada (NSERC) Discovery Grant RGPIN-2019-06467 and Discovery Horizons Grant DH-2025-00119.

\subsection*{Competing Interests}
The authors declare no competing interests.

\subsection*{Author Contributions}
M. Sabouri: conceived the idea, designed the experiments, performed data analysis, developed models, and wrote the manuscript.  
G. Hajianfar: conceived the idea, implemented methods, and revised the manuscript.  
A. Rafiei Sardouei, M. Yazdani: implemented methods and revised the manuscript.  
A. Asadzadeh: performed segmentation and labeling.  
S. Bagheri, M. Arabi, S. R. Zakavi, E. Askari, and A. Aghaee: contributed to data collection.  
S. Wiseman, D. Shahriari, H. Zaidi, and A. Rahmim: discussed research efforts and provided feedback on the study and manuscript.

\subsection*{Data Availability}
Data are available from the corresponding author upon reasonable request.

\subsection*{Code Availability}
The code is available at: \url{https://github.com/MaziarSabouri/Stable-Diffusion-Scintigraphy-Augmentation}.

\subsection*{Ethics Approval}
This study was approved by the Research Ethics Committee of Kashan University of Medical Sciences (Approval Code: IR.KAUMS.NUHEPM.REC.1403.022) and was conducted in accordance with the Declaration of Helsinki.

\subsection*{Consent to Participate}
All necessary institutional approvals and participant consents were obtained.

\subsection*{Consent to Publish}
All authors have reviewed and approved the manuscript for publication.

\bibliographystyle{unsrt}
\bibliography{reference}


\end{document}